# Spin-lattice relaxation of Mn-ions in ZnMnSe/ZnBeSe quantum wells measured under pulsed photoexcitation.

M. K. Kneip, D. R. Yakovlev* and M. Bayer,

*Experimentelle Physik II, Universität Dortmund, D-44221 Dortmund, Germany*

A. A. Maksimov and I. I. Tartakovskii

*Institute of Solid State Physics, Russian Academy of Sciences, 142432 Chernogolovka, Russia*

D. Keller, W. Ossau and L. W. Molenkamp

*Physikalisches Institut der Universität Würzburg, D-97074 Würzburg, Germany*

A. Waag

*Institute of Semiconductor Technology, Braunschweig Technical University, D-38106 Braunschweig, Germany*

**Abstract**    The dynamics of spin-lattice relaxation of the Mn-ions in (Zn,Mn)Se-based diluted-magnetic-semiconductor quantum wells is studied by time-resolved photoluminescence. The spin-lattice relaxation time varies by five orders of magnitude from $10^{-3}$ down to $10^{-8}$ s, when the Mn content increases from 0.4 up to 11%. Free carriers play an important role in this dynamics. Hot carriers with excess kinetic energy contribute to *heating* of the Mn system, while *cooling* of the Mn system occurs in the presence of cold background carriers provided by modulation doping. In a $Zn_{0.89}Mn_{0.11}Se$ quantum well structure, where the spin-lattice relaxation process is considerably shorter than the characteristic lifetime of nonequilibrium phonons, also the phonon dynamics and its contribution to heating of the Mn system are investigated.

PACS:  75.50.Pp,  78.55.Et,  78.20.Ls,  85.75.-d

## 1. Introduction

In recent years the fast developing research area of spintronics has attracted much attention. New device concepts, based on precise manipulation of electron spin rather than on controlling electron charge have been invented [1, 2]. Operation of future spintronics devices inevitably implies experimental possibilities to create spin-polarized electrons (spin injection), to controllably switch their spin state (spin manipulation), and to store orientation and/or phase of the spin state (spin memory, spin coherence). Diluted magnetic semiconductors (DMS) are among the most promising materials for these devices and are widely used nowadays for testing novel design concepts. DMS based on II-VI semiconductors like (Cd,Mn)Te, (Zn,Mn)Se, (Zn,Mn)Te are very suitable for this purpose due to the possibility to use optical spectroscopy in such studies.

Often $Mn^{2+}$ ions form the magnetically active component in II-VI DMS. $Mn^{2+}$ has a total magnetic moment 5/2 based on pure spin momentum provided by five the $d$-shell electrons and substitutes metal ions in the cation sublattice. Strong exchange interaction of the localized $Mn^{2+}$ magnetic moments with conduction band electrons ($s$-$d$ interaction) and/or valence band holes ($p$-$d$ interaction) creates a variety of giant magneto-optical and magneto-transport effects in DMS. Among them are giant Zeeman splittings of the band states that may exceed 100 meV at low temperatures, giant Faraday and Kerr rotation effects, formation of magnetic polarons *etc*. [3, 4]. The magnetic properties of the Mn-ion system, which depend strongly on Mn concentration, play a key role in these effects. For example, neighbouring Mn ions interact anti-ferromagnetically which leads to the formation of high ordered clusters and spin-glass phases at higher Mn concentrations [3]. Also the dynamical properties, namely spin dephasing and spin-lattice relaxation rates, of the localized Mn spins are controlled by concentration dependent exchange interactions between the Mn-ions [5]. It has been shown that the spin-lattice relaxation (SLR) time varies by several orders of magnitude from milliseconds down to nanoseconds with increasing Mn content [5-7].

As the Mn spin-lattice relaxation is of key importance for the dynamical processes in DMS, during last decade a number of studies have been performed for bulk (Cd,Mn)Te [8-12] and for (Cd,Mn)Te-based heterostructures [6, 13, 14] with Mn concentrations not exceeding 5%. The SLR rate is a strong function of the lattice temperature [12], and it decreases weakly in magnetic fields $B$ below 12 T [8-10, 6] but increases rapidly ($t_{SLR}^{-1} \propto B^5$) for $B$>17 T [11]. In DMS heterostructures the SLR rate can be accelerated by presence of free carriers, that can be provided by modulation doping [13, 14], and also by realization of the concept of heteromagnetic nanostructures with varying Mn concentrations across the sample [15].

Only very recently experimental studies have been extended toward (Zn,Mn)Se heterostructures [7, 15]. In this paper we report on a comprehensive investigation of the SLR dynamics of Mn-ions in (Zn,Mn)Se quantum well structures with widely varying parameters. Implementation of pulsed laser excitation for the Mn spin heating allows us to extend range of the dynamics time constants down to nanoseconds, facilitating the study of samples with Mn content covering the large range from 0.4 up to 11%. Further, besides nominally undoped also modulation doped samples have been studied, showing that the presence of free carriers accelerates considerably the SLR rates.

## 2. Experimental Details

### 2.1 Samples

The $Zn_{1-x}Mn_xSe/Zn_{1-y}Be_ySe$ quantum well heterostructures were grown by molecular beam epitaxy on (100)-oriented GaAs substrates. The substrates were covered by a buffer consisting of 10-Å-thick BeTe, 20-Å-thick ZnSe, and 4000-Å-thick $Zn_{0.97}Be_{0.03}Se$ layers, to improve the surface quality and the lattice matching with the barrier material. Subsequently the quantum well (QW) structures were grown. The structural parameters for the studied samples are collected in Table 1. The Mn content in the $Zn_{1-x}Mn_xSe$ QW layers was varied from $x$=0.004 to 0.11. Most of the samples were nominally undoped, so that the background electron density in the wells does not exceed $10^{10}$ cm$^{-2}$. Also a set of modulation doped samples with two-dimensional electron gas concentrations up to $5.5\times10^{11}$ cm$^{-2}$ was grown to study the effect of free carriers on the SLR dynamics. Information about the optical characterization of the structures can be found in Refs. [16, 17].

### 2.2 Experimental technique

A common way to measure the spin-lattice relaxation rate of the Mn-spin system is to heat it by an impact pulse and to follow the evolution of the Mn spin temperature ($T_{Mn}$) while the system is returning to equilibrium. Here we use an optical method to measure $T_{Mn}$ through the giant Zeeman splitting of the excitonic states. The latter can be measured from the spectral position of the excitonic photoluminescence (PL) emission line. This method has been used in combination with the impact of nonequilibrium phonons generated by electrical pulses [6]. Here it will be modified for the use of laser pulses as sources of the dynamic impact for heating the Mn system.

The optical measurements were performed at a bath temperature of 1.6 K for which the samples were immersed in superfluid helium. Magnetic fields up to 7 T were applied parallel to the structure growth axis and to the direction of collected light (Faraday geometry). The emission was analyzed (either right-hand $s^+$ or left-hand $s^-$) regarding its circular polarization. A pulsed

YAG laser operating at wavelengths of either 355 nm (third harmonic) or 532 nm (second harmonic) was used for heating the Mn system. The pulse duration was about 7 ns, the maximum peak power ~1 kW at a repetition rate of up to 10 kHz. Time resolved PL spectra were recorded by a gated charge-couple-device (CCD) camera synchronized with the pulsed laser. The gate signal which could be delayed with respect to the laser pulses provided a temporal resolution better than 2 ns.

A characteristic exciton recombination time in ZnSe-based quantum wells is about 100 ps, i.e. it is much shorter than the duration of the laser pulse. In order to get information about relaxation processes exceeding the laser pulse duration we provide additional illumination of the sample with a *cw* HeCd laser at 325 nm. The excitation density of this laser was kept below 0.1 W/cm$^2$, to minimize heating of the Mn system. The laser excitation spots usually were larger than ~1 mm in diameter and only small central parts (< 100µm in diameter) of these spots were projected on the entrance slit of a 0.5-m monochromator. This allows us to avoid uncertainties caused by spatially inhomogeneous excitation.

## 3. Results and discussion

### 3.1 Twofold dynamic impact of the laser excitation for Mn heating

Laser light that is absorbed in a DMS structure generates photocarriers (electrons and holes) with excess kinetic energy. This energy can be transferred to the Mn-spin system via two channels shown schematically in Fig.1. The first channel is provided by exchange scattering of the free carriers with the magnetic ions enabling a direct energy transfer. The second channel is indirect and involves nonequilibrium phonons generated by the free carriers during energy relaxation. Therefore, laser pulses have a two-fold dynamic impact in heating the Mn system. These two impacts may differ in temporal duration and profile and in heating efficiency. Their relative contributions depend strongly on the DMS structure parameters and the excitation conditions [18-21]. After the impact pulse the Mn system temperature will relax back to that of the lattice within the spin-lattice relaxation time.

Let us consider the characteristics of the carrier and phonon impacts for our experimental conditions. The temporal profile of the laser pulse $I_L(t)$ is shown in Fig.2 by the solid line. It can be well described by a Gaussian profile with a half width at a half maximum of ~7 ns. If photocarriers are generated in the (Zn,Be)Se barriers, they are captured in the (Zn,Mn)Se quantum wells and recombine there through the exciton states. Carrier lifetimes in the studied structures do not exceed 100 ps, which is typical for ZnSe-based quantum wells [22]. They have been measured with a streak-camera after ps-pulsed laser excitation. Details will be published elsewhere. Therefore, taking into account our time resolution, the temporal profile of the carrier impact $I_c(t)$ should coincide with the laser pulse $I_L(t)$. This is confirmed in experiment where no distinct difference is observed between the laser excitation dynamics and that the excitonic emission, shown exemplarily for sample #11 ($x$=0.11) by the dots in Fig.2. However, the temporal evolution of the phonon impact $I_{ph}(t)$ differs from the laser pulse. Nonequilibrium phonons are generated by the dissipation of the kinetic energy of the photocarriers. Therefore, while the leading edge of $I_{ph}(t)$ does not exceed ~10 ns (i.e. the integral of the laser pulse), but the trailing edge is determined by the characteristic lifetime of the acoustic phonons in crystals, which is in the order of 1 µs at low temperatures [23]. The temporal profiles of the dynamic impacts of carriers and phonons are given schematically in the inset of Fig.2.

*3.2  Response of the Mn system to the dynamic impacts*

The dynamical response of the Mn temperature $T_{Mn}(t)$ on an impact will differ for $I_c(t)$ and $I_{ph}(t)$, as it is determined by the difference in characteristic times during which energy can be transferred from the carriers to the Mn ions ($\boldsymbol{t}_{c-Mn}$), from the phonons to the Mn ions ($\boldsymbol{t}_{SLR}$) and from the Mn ions back to the lattice ($\boldsymbol{t}_{SLR}$). Therefore the response allows us to measure these times experimentally. However the different impact contributions need to be distinguished from the magnetization relaxation, which is not always trivial. In this work we concentrate solely on the spin-lattice relaxation dynamics in (Zn,Mn)Se. A detailed consideration of the interplay of the different energy transfer scenarios determining $T_{Mn}(t)$ for our double dynamic impact situation is out of the frame of this paper. Here we consider only regimes, which have been realized in our experimental conditions.

When $\boldsymbol{t}_{SLR}$ exceeds the durations of the impact pulses, $\Delta t_c$ for the carrier impact and $\Delta t_{ph}$ for the phonon impact, the SLR time can be measured from the decay of $T_{Mn}(t)$. This regime has been realized experimentally by injection on nonequilibrium phonons, enabling the measurement of $\boldsymbol{t}_{SLR}$ in (Cd,Mn)Te QWs with $x$<0.035 [6]. The situation becomes more complicated when the SLR dynamics is faster than the phonon impact. In Fig.3 we analyze the case $\Delta t_c < \boldsymbol{t}_{SLR} < \Delta t_{ph}$ for different relative contributions of carriers and phonons. The impact profiles are shown by solid lines and the expected $T_{Mn}(t)$ are given by dashed lines. Cases (a) and (b) are for single impact conditions, when one of the contributions strongly dominates over the other. For carrier impact only [case (a)], $\Delta t_c < \boldsymbol{t}_{SLR}$ and $\boldsymbol{t}_{SLR}$ determines the decrease of $T_{Mn}(t)$ toward the lattice temperature. In case (b) $\boldsymbol{t}_{SLR} < \Delta t_{ph}$ and SLR time can be measured from the rise of $T_{Mn}(t)$. This is possible due to the sharp rise of the phonon impact $I_{ph}(t)$, which is $\sim\Delta t_c$. The decrease of $T_{Mn}(t)$ follows the slow decay of the phonon impact. The double impact case (c) is realized for the condition $\Theta_c > \Theta_{ph}$, where $\Theta_c$ and $\Theta_{ph}$ are the maximum $T_{Mn}$ that are obtained under carrier and phonon impacts, respectively. $\Theta_c$ and $\Theta_{ph}$ can be used to compare the efficiencies of the Mn system heating by the different impacts. In case (c), the decrease of $T_{Mn}(t)$ has fast and slow components, corresponding to the SLR and the phonon impact, respectively.

It should be noted that Fig. 3(a) qualitatively describes as well the situation when the SLR time is longer than the duration of both dynamic impacts, i.e. $\Delta t_c, \Delta t_{ph} < t_{SLR}$. In this case, as already found above, the SLR time can be determined from the decay of $T_{Mn}(t)$.

### 3.3 Optical detection of the magnetization dynamics

Here we are interested in the dynamical response of the magnetization due to the impact of a laser pulse. To study this impact, a finite equilibrium magnetization has to be induced by application of an external magnetic field and its time evolution has to be detected. Optical spectroscopy offers very sensitive methods for such a measurement. In the present studies we exploit the internal Mn-spin thermometer, which is provided by the high sensitivity of the giant Zeeman splitting of excitons (band states) to the polarization of the magnetic ions.

The giant Zeeman splitting is proportional to the magnetization and thus to the average spin of the Mn ions.

$$\Delta E_Z = \left( \boldsymbol{d}_e \boldsymbol{a} - \boldsymbol{d}_h \boldsymbol{b} \right) N_0 x \langle S_z \rangle \qquad (1)$$

Here $N_0 \boldsymbol{a} = 0.26$ eV and $N_0 \boldsymbol{b} = -1.31$ eV are the exchange constants in $Zn_{1-x}Mn_xSe$ for conduction and valence band, respectively [24]. $N_0$ is the inverse unit-cell volume and $x$ is the Mn mole fraction. The parameters $\boldsymbol{d}_e$ and $\boldsymbol{d}_h$ have been introduced to account for the leakage of the electron and hole wave functions into the nonmagnetic ZnBeSe barriers. For the studied structures the parameters are very close to unity as the carrier wave functions are well localized in the DMS quantum wells. $\langle S_z \rangle$ is the mean thermal value of the Mn spin component along the magnetic field $B = B_z$ at a Mn spin temperature $T_{Mn}$. It can be expressed by the the Brillouin function $B_{5/2}$ as

$$\langle S_z \rangle = -S_{eff}(x)\, B_{5/2}\left[\frac{5 g_{Mn}\boldsymbol{m}_B B}{2k_B\left(T_{Mn}+T_0(x)\right)}\right]. \qquad (2)$$

Here $g_{Mn} = 2$ is the g-factor of the $Mn^{2+}$ ions. $S_{eff}$ is the effective spin and $T_0$ is the effective temperature. These parameters permit a phenomenological description of the antiferromagnetic Mn-Mn exchange interaction. For their values refer e.g. to Fig. 4 of Ref. [17].

Photoluminescence spectra of the $Zn_{0.89}Mn_{0.11}Se/Zn_{0.89}Be_{0.11}Se$ DMS quantum well structure (sample #11) are given in Fig.4a. The two bottom spectra were detected under *cw* laser illumination with very low excitation density to avoid heating of the Mn system above the bath temperature. The giant Zeeman shift of the emission line amounts to about 40 meV at *B*=3 T. The three upper spectra show the emission line at different delays in respect to the impact laser pulse. Just after the pulse ($\Delta t \approx 10$ ns) the Zeeman shift is reduced to ~14 meV, which corresponds to a heating of the Mn system up to $T_{Mn} = 17$ K. After one microsecond the line is shifted back to lower energies, reflecting the cooling of the Mn system.

There are two characteristics of the magneto-optical spectra, which can be exploited for obtaining information about the temperature of the Mn-spin system. Both are related to the giant Zeeman splitting effect of the conduction and valence band states. The first characteristic is the energy shift of the emission line. It is convenient to use it in relatively strong magnetic fields (exceeding ~0.5 T), for which the Zeeman shift can be clearly detected. For measurements in weak magnetic fields (below ~0.5 T) the circular polarization degree of emission can be analyzed. It has been shown that both characteristics provide the same information about the Mn spin temperature (see Fig. 10 in Ref. [17]).

Figure 4 (b) shows the time evolution of both Zeeman shift and polarization degree induced by laser pulse excitation in a $Zn_{0.89}Mn_{0.11}Se/Zn_{0.89}Be_{0.11}Se$ QW. Fast heating of the Mn ions occurs during action of the laser pulse and is reflected by a high-energy shift of the PL maximum by 23 meV and a decrease of the polarization degree from 0.9 to 0.2. After the laser pulse the Mn system temperature relaxes towards equilibrium with a relaxation time constant of ~23 ns. However, it saturates at a level which exceeds the bath temperature. We will show below that this level is controlled by nonequilibrium phonons. To reach the equilibrium temperature of 1.6 K takes much longer time of a few µs. Remarkably the Zeeman shift and the polarization degree show very similar temporal behaviors and therefore both of them are well suited for optical detection of the spin-lattice dynamics. In the following we will limit ourselves to a Zeeman shift analysis to measure the SLR time in (Zn,Mn)Se structures with different Mn contents.

### 3.4 Low Mn concentration: SLR times longer then the impact duration

For low Mn concentrations, where the SLR dynamics takes considerably longer than the characteristic time of the phonon impact of ~1 µs, the SLR times can be extracted from the decay of the dynamical response (as sketched in Fig.3 (a) for $\Delta t_c, \Delta t_{ph} < t_{SLR}$). In Fig.5 the spectral shift of PL line $\Delta E_{PL}$ induced by the laser impact is plotted as a function of time for samples with different Mn content. Once more we emphasize that the measured values for $\Delta E_{PL}$ are proportional to the changes in magnetization, i.e. in the average spin of the Mn ions. For convenient comparison of the different samples the data have been normalized by the maximum shift $\Delta E_{PL}^{max}$ and have been plotted on a logarithmic scale. As seen from the linear dependence on this scale, the spin relaxation can be well described by a single exponential behaviour. This holds for all samples with Mn content ranging from 0.004 up to 0.035, but the increase on Mn concentration is accompanied by a strong decrease of the SLR times from 960 to 11 µs.

In Fig.6 the spin-lattice relaxation time is plotted as a function of Mn content. The full circles show the measured data for (Zn,Mn)Se-based structures, while the open symbols give the literature data for (Cd,Mn)Te. The (Zn,Mn)Se values for $t_{SLR}$ follow the same trend as those for (Cd,Mn)Te and further support the idea that the relaxation process is provided by anisotropic spin interactions of the magnetic ions that are rather insensitive to the ion host material. For $x>0.04$ the SLR dynamics approaches the duration of the phonon impact. The modifications of the dynamical response in this case are considered in the next section.

### 3.5 High Mn concentration: SLR times comparable with the impact duration

As discussed in Sec. 3.2, the double impact nature of the laser excitation leads to a complicated dynamical response of the magnetic ion system under conditions when the SLR time is shorter than the phonon impact time. This regime $\Delta t_c < t_{SLR} < \Delta t_{ph}$ corresponds to the situation sketched in Fig.3 (c). Experimentally it is realized for the $Zn_{0.89}Mn_{0.11}Se/Zn_{0.94}Be_{0.06}Se$ structure available here. The Mn temperature obtained from the spectroscopic data at $B=3$ T are shown in the two panels of Fig.7. Two different laser excitation wavelengths were used in these studies.

Laser excitation with 532 nm wavelength (photon energy 2.33 eV) is not absorbed by the (Zn,Mn)Se QW and also not by the (Zn,Be)Se barriers, and therefore does not generate carriers in the QWs. However, it is absorbed in GaAs and the Mn system in the (Zn,Mn)Se QW is heated by the phonon impact arising from phonon emission during relaxation only. The phonons are generated in GaAs at distances of less than 1 µm from the (Zn,Mn)Se QW and thus the delay of phonon impact is less than ~1 ns, which is negligible for our particular experimental situation.

The SLR time in this case controls the rise of the $T_{Mn}(t)$ signal [case (b) in Fig.3]. The dynamics is given by circles in Fig.7 (a) from which a $t_{SLR} \approx 25$ ns has been extracted from a fit to the data, shown by the solid line. The decay of this signal with a time constant of about 0.6 µs is due to the phonon dynamics, see Fig.7 (b).

355 nm wavelength excitation (photon energy 3.49 eV) leads to absorption in the immediate region of the II-VI heterostructure and should cause a double impact. The dynamics of the Mn system shown by triangles in the Fig.7 (a) follows the scenario of case (c) in Fig.3. The carrier impact drives up the Mn temperature to $\Theta_c \approx 13$ K during the 10 ns of laser pulse action. Afterwards the Mn system relaxes on a time scale of $t_{SLR} \approx 25$ ns to $\Theta_{ph} \approx 4.2$ K, which is controlled by the phonon impact. At delay times longer than ~100 ns $T_{Mn}(t)$ follows the phonon impact $I_{ph}(t)$ as can be seen from Fig. 7 (b) which gives the data on a logarithmic scale to show a wide time range.

An independent confirmation to our assignment of the dynamical ranges is delivered by the power dependence (inset of Fig.7 (b)). The SLR time decreases from 70 down to 20 ns with increasing power. This is in accord with the known trend of the shorter SLR times at higher lattice temperatures [6, 10]. Simultaneously the time characterizing the phonon dynamics increases from 350 to 1200 ns, which is due to a strong decrease of the mean free path of non-equilibrium phonons with increase of their average frequency [25]. Thus, under higher optical excitation, the propagation of phonons is hindered and becomes slower [25, 26], resulting in longer lifetimes of non-equilibrium phonons inside the sample [23].

The values of $t_{SLR}$ obtained for this sample at the lowest and highest power densities are shown by the two data points in Fig.6. It should be noted that $t_{SLR}$ for sample #10 ($x$=0.1) was measured under high excitation density of $P \approx 100$ kW/cm$^2$, and thus corresponds to a spin-lattice relaxation time at a temperature higher than the bath temperature. Its expected value shift to the regime of low excitation density is shown by the arrow.

*3.6   Effect of free carriers on SLR.*

The dynamics of the spin-lattice relaxation can be modified significantly by the presence of free carriers, which are strongly coupled with both the magnetic ions and the phonons. Therefore the free carriers may serve as a bypass channel for the slow direct spin-lattice relaxation (see scheme in Fig.8). An acceleration of SLR has been experimentally found in (Cd,Mn)Te QWs modulation doped with either electrons or holes [13, 14].

In order to study this effect in (Zn,Mn)Se QWs we examine the set of samples #1–#4 with $x$=0.004 and electron densities varying from $10^{10}$ (nominally undoped) up to $5.5 \times 10^{11}$ cm$^{-2}$. The measured SLR times in Fig. 8 show a decrease from 960 down to 70 µs, which means that in the doped samples $t_{SLR}$ is 14 times shorter than in the undoped case, confirming that the SLR dynamics is dominated by the free carriers.

## 4. Conclusions

Pulsed laser excitation is a very reliable tool to measure spin-lattice relaxation times in DMS heterostructures. Its implementation allows us to extend the dynamical range for measurements of the SLR times to cover magnetic ion concentrations from 0.4 up to 11% in (Zn,Mn)Se-based heterostructures. The SLR relaxation times vary by five orders of magnitude from $10^{-3}$ down to $10^{-8}$ seconds. The times and their dependence on Mn content are very similar in (Zn,Mn)Se and (Cd,Mn)Te DMS materials. This confirms that the SLR dynamics is controlled by interactions within the system of Mn ions, but is only weakly sensitive to the II-VI semiconductor host material, into which the Mn ions are embedded.

Presence of free carriers is an important influence factor for the spin dynamics of the magnetic ions. The dominating implication of the laser pulse on the Mn heating comes from hot photocarriers transferred their excess kinetic energy. Also the presence of a two-dimensional electron gas in modulation doped samples may accelerate the SLR dynamics by more than an order of magnitude, offering a way to control SLR dynamics by applying an electric field through a gate.

**Acknowledgments**

We appreciate fruitful discussions with A. V. Akimov and V. D. Kulakovskii. We are thankful to the Roper Scientific GmbH, Germany for lending us the PIMAX gated CCD camera used for the time-resolved experiments. This work has been supported by the BMBF nanoquit program, the DARPA QuIST program, and the Deutsche Forschungsgemeinschaft with the research stays in Dortmund of A.A. Maksimov (grants Nos. DFG 436RUS17/81/04, DFG 436 RUS 17/49/05) and I.I. Tartakovskii (grants Nos. DFG 436 RUS 17/80/03, DFG 436 RUS 17/97/04) as well as within the Sonderforschungsbereich 410. Also support by the Russian Foundation for Basic Research (grants Nos. 05-02-17288, 04-02-16852), and by INTAS (grant No. 03-51-5266) is acknowledged.

**Table caption.**

Table 1.   Technological parameters and experimental values for (Zn,Mn)Se based samples studied in this paper.

Table 1.

| sample no. | code | Mn content, $x$ | Electron concentration $\times 10^{10}$ [cm$^{-2}$] | SLR time [μs] | QW width [Å] | barrier width [Å] | Be content, $y$ | number of periods | PL linewidth, FWHM [meV] |
|---|---|---|---|---|---|---|---|---|---|
| 1 | CB1542 | 0.004 | undoped | 960 | 100 | 200 | 0.06 | 5 | 2.0 |
| 2 | CB2033 | 0.004 | 3 | 550 | 100 | --- | 0.06 | 1 | 2.0 |
| 3 | CB2034 | 0.004 | 32 | 100 | 100 | --- | 0.06 | 1 | 4.2 |
| 4 | CB2037 | 0.004 | 55 | 70 | 100 | --- | 0.06 | 1 | 6.9 |
| 5 | CB1651 | 0.012 | undoped | 600 | 150 | --- | 0.06 | 1 | 2.2 |
| 6 | CB1581 | 0.015 | undoped | 530 | 150 | --- | 0.04 | 1 | 2.1 |
| 7 | CB2422 | 0.02 | undoped | 100 | 100 | --- | 0.06 | 1 | 2.6 |
| 8 | CB2169 | 0.035 | undoped | 11 | 100 | --- | 0.06 | 1 | 2.3 |
| 9 | CB1433 | 0.06 | undoped | 1.8 | 200 | 100 | 0.05 | 10 | 2.7 |
| 10 | CB1340 | 0.1 | undoped | 0.03 | 3000 | --- | --- | epilayer | 8.5 |
| 11 | CB886 | 0.11 | undoped | 0.02-0.07 | 100 | 95 | 0.11 | 10 | 6.5 |

**Figure captions**

Fig.1    Channels for energy transfer in the process of heating (double-line arrows) and cooling (single-line arrow) of Mn spin system under laser excitation. The dashed-line arrow shows phonon generation due to energy relaxation of photocarriers. Laser light heats the Mn system by carrier $I_c(t)$ and phonon $I_{ph}(t)$ impacts. The system relax towards equilibrium (given by the lattice temperature) with the spin-lattice relaxation time $t_{SLR}$.

Fig.2    Temporal evolution of a Nd:YAG laser pulse (solid line) and of the photoluminescence signal of a $Zn_{0.89}Mn_{0.11}Se/Zn_{0.89}Be_{0.11}Se$ QW sample (dots). In the inset the two impacts to the Mn heating are shown schematically: $I_c(t)$ is the *carrier impact* due to direct energy transfer from carriers to the ions and $I_{ph}(t)$ is the *phonon impact* due to indirect energy transfer mediated by nonequilibrium phonons.

Fig.3    Dynamical response of the Mn system on the impact pulses under various experimental conditions: (a) carrier impact only with a duration shorter than $t_{SLR}$; (b) phonon impact only with duration longer than $t_{SLR}$; (c) double impact of carriers and phonons with $\Delta t_c < t_{SLR} < \Delta t_{ph}$ and $\Theta_c > \Theta_{ph}$.

Fig.4    (a) PL spectra of a $Zn_{0.89}Mn_{0.11}Se/Zn_{0.89}Be_{0.11}Se$ sample. The two bottom spectra are taken under very low *cw*-laser excitation, the others are recorded at different time delays $\Delta t$ with respect to the YAG laser pulse maximum: $\Delta t \approx 0$ ns (corresponds to the pulse maximum), 10 ns (just after the pulse), and 1 µs later. The excitation density $P \approx 54$ kW/cm$^2$, magnetic fields $B = 3$ T (solid lines) and $B = 0$ (dashed line), bath temperature $T = 1.6$ K.

(b) Temporal variation of the PL spectral line position $E_{PL}$ at a magnetic field $B = 3$ T (closed circles) and of the circular polarization degree $P_c$ at $B = 0.12$ T (open circles) in a $Zn_{0.89}Mn_{0.11}Se/Zn_{0.89}Be_{0.11}Se$ QW structure. The excitation density $P \approx 12$ kW/cm$^2$, bath temperature $T = 1.6$ K. The laser pulse maximum position is indicated by the vertical arrow.

Fig.5    Temporal evolution of PL spectral line shift $\Delta E_{PL}$ which is proportional to the change $\Delta M$ of the Mn magnetization in $Zn_{1-x}Mn_xSe/Zn_{0.94}Be_{0.06}Se$ QWs with different Mn content $x$. The spin-lattice relaxation time $t_{SLR}$ decreases from 960 to 11 μs with increasing $x$ from 0.004 to 0.035. $B = 3$ T.

Fig.6    Spin-lattice relaxation time as a function of Mn content for nominally undoped $Zn_{1-x}Mn_xSe/Zn_yBe_{1-y}Se$ structures (closed circles). Values are also listed in Table 1. Open symbols represent the literature data for $Cd_{1-x}Mn_xTe$ bulk samples (triangles and circles) [12, 10] and $Cd_{1-x}Mn_xTe$-based heterostructures (diamonds) [6]. The dashed lines indicate typical lifetimes of nonequilibrium phonons.

Fig.7    Dynamics of the Mn temperature $T_{Mn}$ under 355 nm ($P \approx 9$ kW/cm$^2$, double impact by carriers and phonons) and 532 nm ($P \approx 15$ kW/cm$^2$, single impact by phonons only) laser excitation in $Zn_{0.89}Mn_{0.11}Se/Zn_{0.89}Be_{0.11}Se$ QWs. Experimental data are given by triangles and circles, respectively. The solid lines show fits of the SLR dynamics with $t_{SLR} = 25$ ns. The bath temperature of 1.6 K is indicated by the horizontal dashed line. The vertical dashed line indicates the maximum of laser pulse at 50 ns. Note the logarithmic time scale in panel (b). The *inset* shows spin-lattice and nonequilibrium phonon relaxation times measured for different powers of 355 nm laser excitation. $B = 3$ T.

Fig.8　Dependence of the spin-lattice relaxation time $t_{SLR}$ on the concentration of free electrons $n_e$ in $Zn_{0.996}Mn_{0.004}Se/Zn_{0.94}Be_{0.06}Se$ QWs. The solid line is a guide to the eye. Scheme in the *inset* illustrates the bypass channel for energy transfer from the Mn system to the lattice through the two-dimensional electron gas (2DEG).

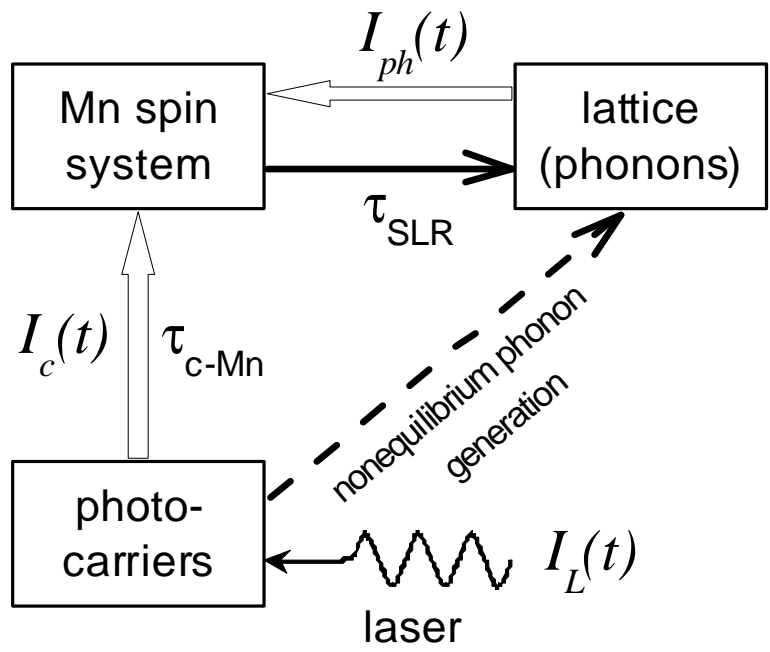

Figure 1.

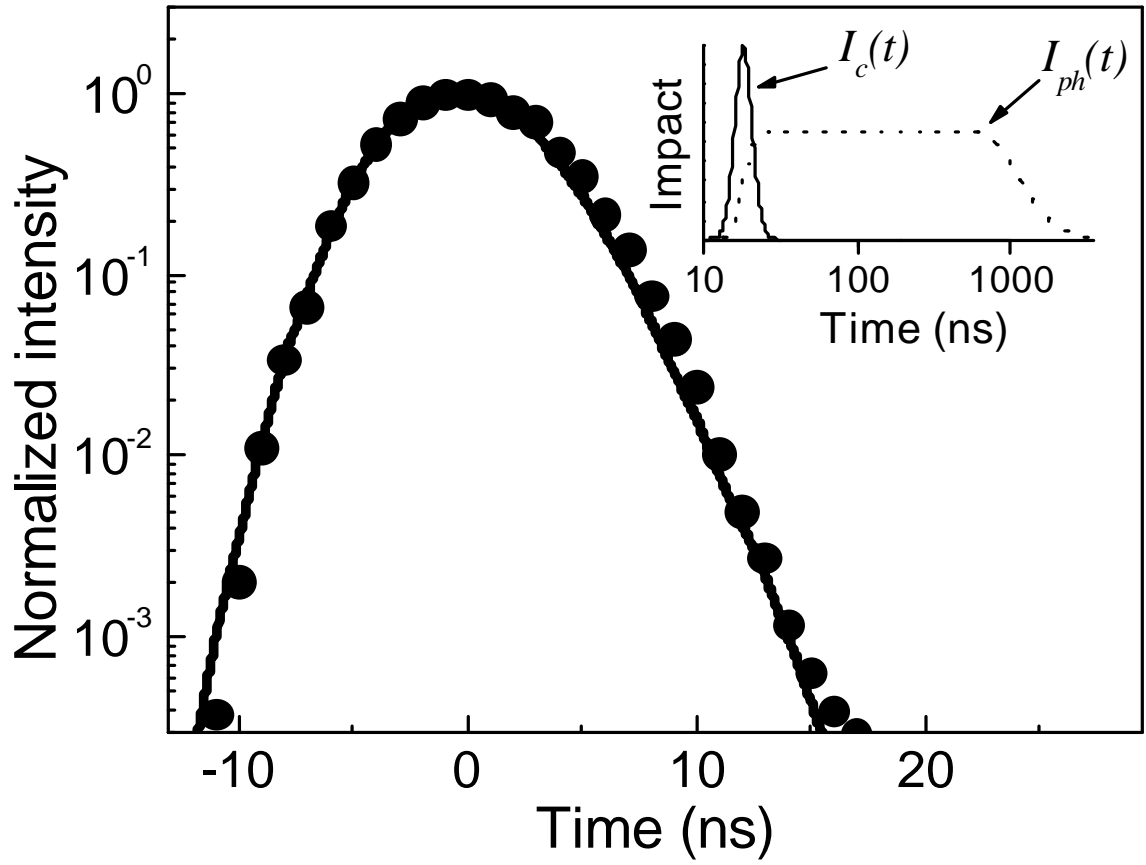

Figure 2.

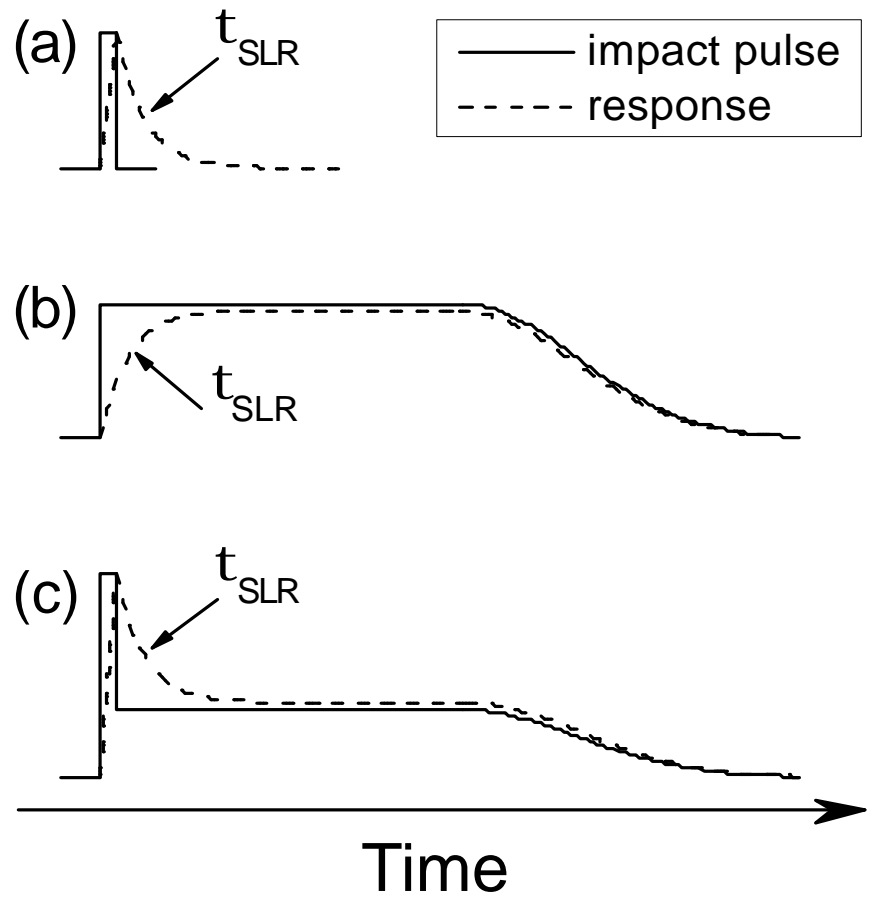

Figure 3.

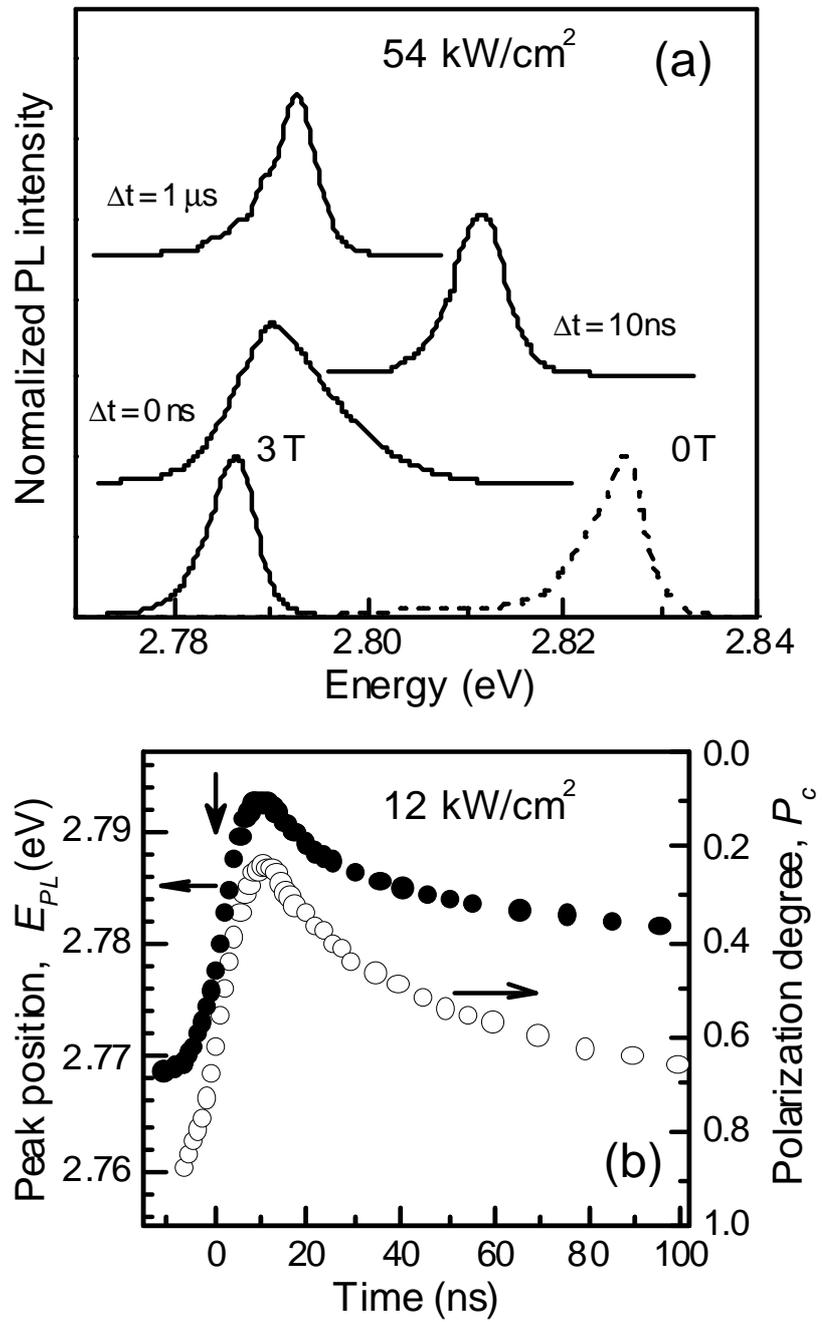

Figure 4.

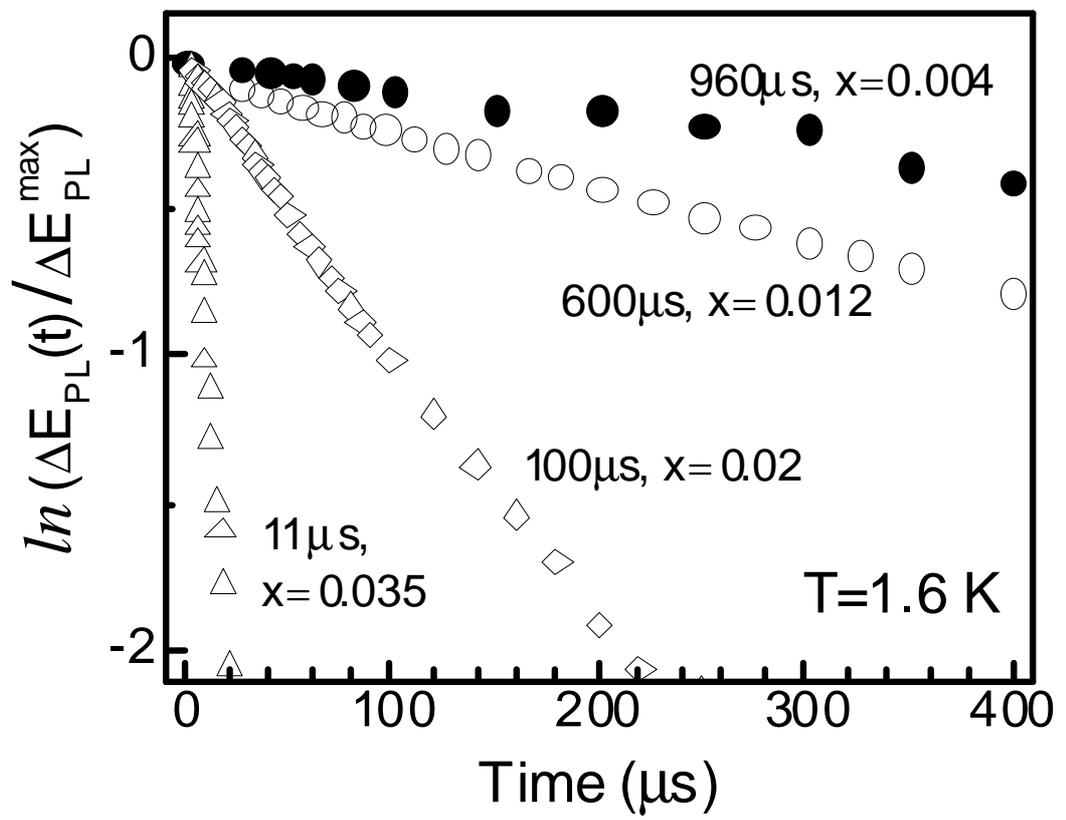

Figure 5.

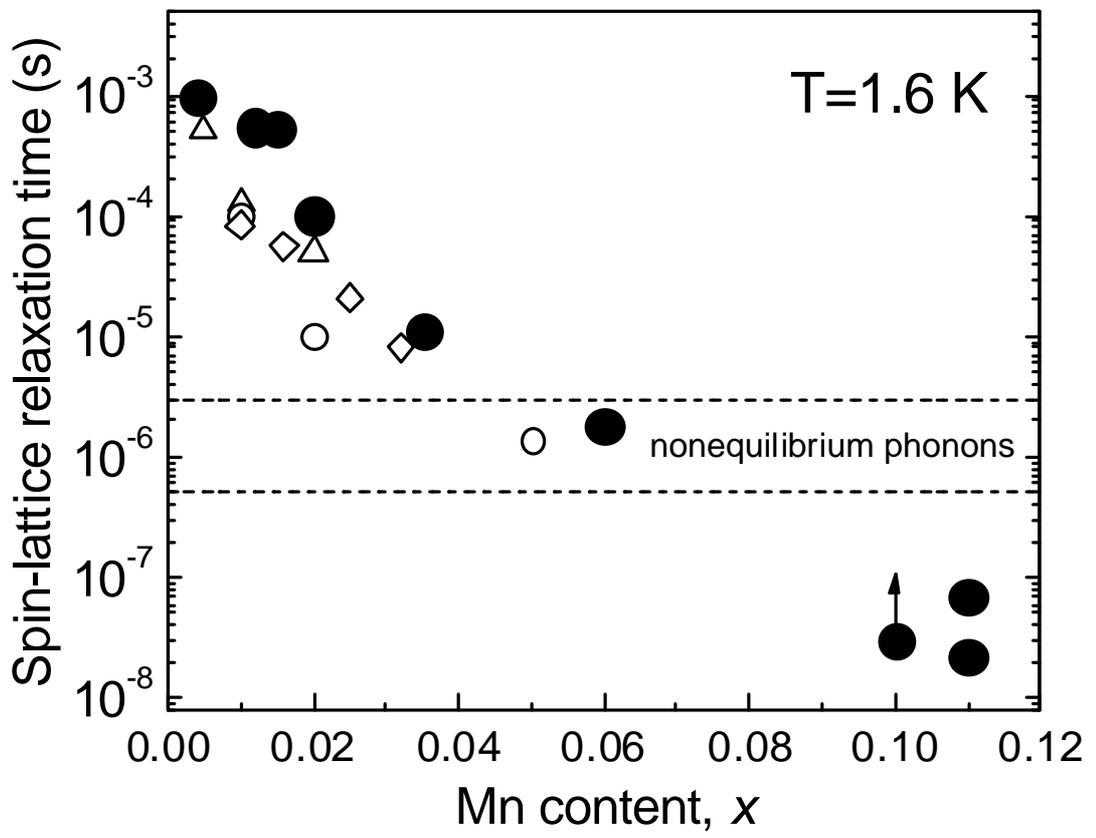

Figure 6.

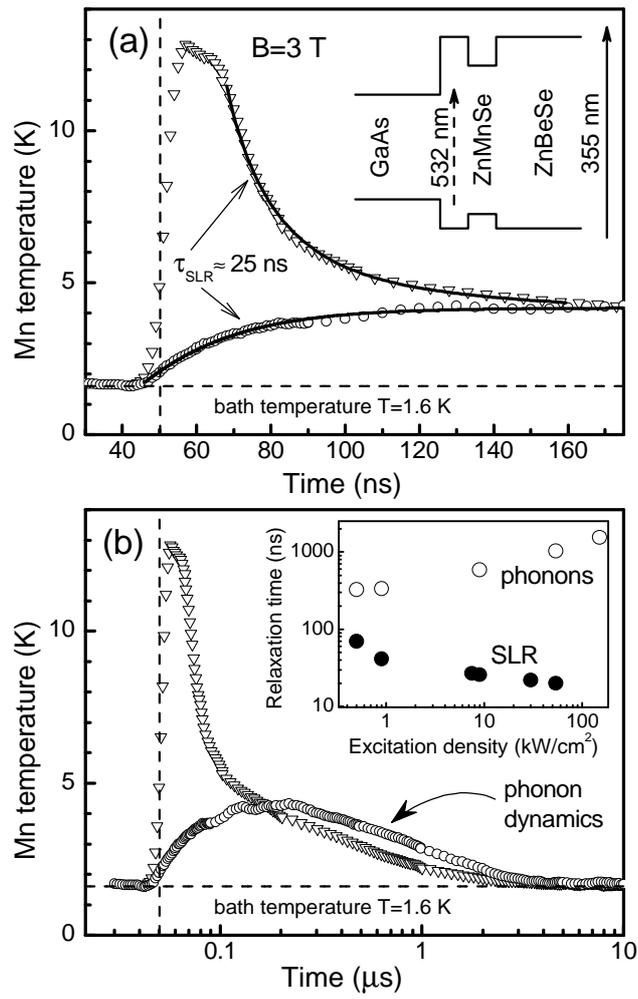

Figure 7.

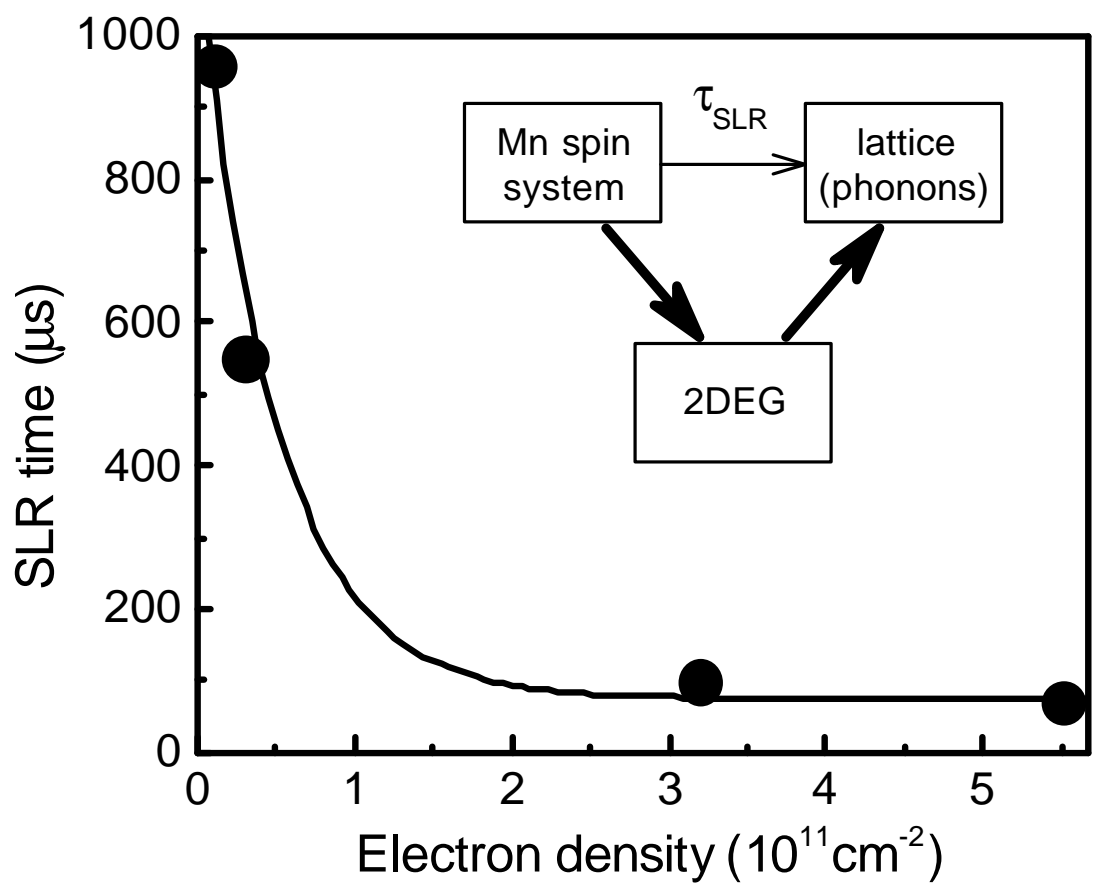

Figure 8.